%% file: text_new.tex
\documentclass[oldversion,prints]{aa}
\usepackage{supertabular}
\usepackage{longtable}
\usepackage{graphicx}
\titlerunning{Canes Venatici I cloud of galaxies seen in the $H\alpha$ line}

\begin{document}
\title{Canes Venatici I cloud of galaxies seen in the $H\alpha$ line}
\author{S.S. Kaisin
\and I.D. Karachentsev}
\authorrunning{Kaisin \& Karachentsev}

\institute{Special Astrophysical Observatory, Russian Academy
	  of Sciences, N.\ Arkhyz, KChR, 369167, Russia}
\abstract{
  We present results of $H\alpha$ imaging for 42 galaxies in the nearby
low-density cloud Canes Venatici I populated mainly by late-type objects.
Estimates of the $H\alpha$ flux and integrated star formation rate ($SFR$)
are now available for all 78 known members of this scattered system,
spanning a large range in luminosity, surface brightness, $HI$ content and
$SFR$. Distributions of the CVnI galaxies versus their $SFR$, blue absolute
magnitude and total hydrogen mass are given in comparison with those
for a population of the nearby virialized group around M81. We found
no essential correlation between star formation activity in a galaxy and
its density environment. A bulk of CVnI galaxies had enough time to
generate their baryon mass with the observed $SFR$. Most of them possess
also a supply of gas sufficient to maintain their observed $SFR$s
during the next Hubble time.  }

\maketitle
\section{Introduction}
The distribution over the sky of 500 galaxies of the Local volume  with
distances  within 10 Mpc shows considerable inhomogeneities due to the presence of
groups and voids. Apart from several virialized groups, like the group around
the galaxy M81, an amorphous association of nearby galaxies in the Canes
Venatici constellation was noted by many authors (Karachentsev 1966, de
Vaucouleurs 1975, Vennik 1984, Tully 1988). The boundaries of it are rather
uncertain. Roughly, in a circle of radius $\sim20\degr$ around the galaxy
NGC~4736 there are about 80 known galaxies with $D<10$ Mpc, which corresponds
to a density contrast on the sky $\Delta N/N\sim4$. Inside of this complex
the distribution of galaxies is also inhomogeneous, showing some clumps
differing in their location on the sky and distances in depth. About 70\%
of the population of the cloud accounts for irregular dwarf galaxies, whose
masses are obviously insufficient to keep such a system in the state of
virial equilibrium. As data on distances of galaxies accumulated, a
possibility appeared of studying the kinematics of the cloud in details.
As was shown by Karachentsev et al. (2003), the CVnI cloud is in a state
close to the free Hubble expansion, having a characteristic crossing
time of about 15 Gyr.

Being a scattered system with rare interactions between galaxies, the nearby
CVnI cloud is a unique laboratory for studying star formation processes in
galaxies running independently, without a noticeable external influence.
Kennicutt et al. (1989), Hoopes et al. (1999), van Zee (2000), Gil de
Paz et al. (2003), James et al. (2004) and Hunter \& Elmegreen (2004)
conducted observations in the $H\alpha$ line of three dozen galaxies
of this complex, which made it possible to determine the star formation
rate ($SFR$) in them. However, more than half of other members of the cloud
proved to be out of vision of these authors. Our task consisted in the
completion of  $H\alpha$-survey of the population of the CVnI cloud.
The results of our observations and their primary analysis are presented
in this paper.

\section{Observations and data reduction}
CCD images in the $H\alpha$-line and continuum were obtained for 42 galaxies
of the CVnI cloud during observing runs from 2001 March to 2006 May. An
average seeing was $1.8\arcsec$. All the observations were performed in
the Special Astrophysical Observatory of the Russian Academy of Sciences with
the BTA 6-m telescope equipped with the SCORPIO focal reducer (Afanasiev et
al.\ 2005). A CCD chip of 2048$\times$2048 pixels provides a total field of
view of about 6.1$\arcmin$ with a scale of 0.18$\arcsec$/pixel. The images
in $H\alpha$+[NII] and continuum were obtained via observing the galaxies
though a narrow-band interference filter $H\alpha (\Delta\lambda=$75\AA)
with an effective wavelength $\lambda$=6555\AA\ and two medium-band
filters for the continuous spectrum SED607 with
$\Delta\lambda$=167\AA, $\lambda$=6063\AA\, and SED707 with
$\Delta\lambda$=207\AA, $\lambda$=7063\AA,\, respectively. Typical exposure
times for the galaxies were $2\times300$s in the continuum and
$2\times600$s in $H\alpha$. Since the range of radial velocities
in our sample is small, we used one and the same $H\alpha$ filter
for all the observed objects. Table 1 presents a brief synopsis of these
runs, where columns are: (1) the galaxy name, (2) the date of observation,
(3) the total exposure time in seconds; a colon means that the sky
was not photometric.
Our data reduction followed the standard practice and was performed within
the MIDAS package. For all the data bias was subtracted and the
images were flat-fielded by twilight flats. Cosmic particles were
removed and the sky background was subtracted.
The next operation was to bring all the images of a given
object into coincidence. Then the images in the continuum were
normalized to $H\alpha$ images using 5--15 field stars and subtracted.
$H\alpha$ fluxes were obtained for the continuum-subtracted images,
using spectrophotometric standard stars from Oke (1990)
observed in the same nights as the objects.
The investigation of measurement errors contributed from the
continuum subtraction, flat-fielding and scatter in the zeropoints
has shown that they have typical values within 10\%. We did not
correct $H\alpha$ fluxes for the contribution of the [NII] lines,
because it is likely to be small for the majority of low-luminosity galaxies
in our sample.
\section{Results}
The images of 42 galaxies that we obtained in the CVnI
cloud are displayed in the form of mosaic in Fig.1.
The left and right image of every galaxy corresponds to the sum and
difference of the frames exposed in $H\alpha$ and continuum.
The frame size is about $4\arcmin\times4\arcmin$, the North and
East directions are indicated by the arrows, the scale is denoted in
the lower right angle.

Some basic properties of 78 galaxies located in the CVnI cloud are listed
in Table 2. It also includes the data on the galaxies that were observed
by other authors earlier. The columns of Table 2 contain the following
characteristics of the cloud members taken mainly from the Catalog of
Neighboring Galaxies (CNG); Karachensev et al. 2004: (1) the galaxy name,
(2) and (3) the equatorial coordinates for the epoch J2000.0,
(4)  morphological type in numerical code according to
de Vaucouleurs et al. (1991), (5) the tidal index ($TI$) following from the CNG;
i.e. for every galaxy ``$i$'' we have found its ``main disturber''(=MD),
producing the highest tidal action
$TI_i = \max  \{\log(M_k/D_{ik}^3)\} + C,\;\;\;(k = 1, 2...  N)$,
where  $M_k$ is the total mass of any neighboring potential MD galaxy
(proportional to its luminosity with $M/L_B = 10 M_{\odot}/L_{\odot}$)
separated from the considered galaxy by a space distance $D_{ik}$;
the value of the constant $C$  is chosen so that $TI = 0$, when the
Keplerian cyclic period of the galaxy with respect to its MD equals
the cosmic Hubble time, $T_{0}$; therefore positive values correspond
to galaxies in groups, while the negative ones correspond to field
galaxies. Column (6) gives the galaxy radial velocity (in km s$^{-1}$)
with respect to
the Local Group centroid with the apex parameters adopted in the
NASA/IPAC Extragalactic Database (NED). Column (7) gives the distance
to a galaxy in megaparsecs with allowance made for new measurements
(Karachentsev et al.\ 2006, Tully et al. 2007). Column (8) presents the blue
absolute magnitude of a galaxy with the given distance after correction
for the Galactic extinction $A_b$ from Schlegel et al. (1998) and the
internal absorption in the galaxy determined as
$A_{int} = [1.6 + 2.8(\log V_m-2.2)]\cdot \log (a/b),$
if $V_m > 42.7$ km s$^{-1}$,  otherwise, $A_{int}=0$. Here, $V_m$ is the
rotation velocity of the galaxy corrected for the inclination, and  $a/b$
is the galaxy axial ratio. Therefore, we assume the internal absorption
to be dependent  not only on the inclination, but also on the galaxy
luminosity (Verheijen, 2001).
Column (9) gives  the logarithm of the hydrogen mass of a galaxy,
$\log(M_{HI}/M_{\odot})= \log F_{HI}+2\lg D_{Mpc}+5.37$,
defined from its flux $ F_{HI}$ in the 21 cm line; in some dwarf
spheroidal galaxies the upper limit of the flux was estimated from the
observations by  Huchtmeier et al. (2000). Column (10) gives
the logarithm of the observed
integral flux of a galaxy in the $H\alpha$ +[NII] lines expressed in
terms of  erg cm$^{-2}$sec$^{-1}$. Notes indicate data sources of $SFR$s
according to other authors.
Column (11) gives the star
formation rate in the galaxy on a logarithmic scale,
$SFR(M_{\odot}$/year) = 1.27$\cdot 10^9 F_c(H{\alpha})\cdot D^2$
(Gallagher et al.\ 1984), where the integral flux in the $H\alpha$
line is corrected for the Galactic and internal
extinction as $A(H\alpha) = 0.538\cdot A_b$,
while the galaxy distance is expressed in Mpc.
Columns (12) and (13) give the dimensionless parameters
$p_*=\log([SFR]\cdot T_0/L_B)$ and
$f_*=\log(M_{HI}/[SFR]\cdot T_0)$,
which characterize the past and the future of the process of star formation;
here $L_B$ denotes the total blue luminosity of galaxy in units of solar
luminosity, while $T_0$ is the age of the universe assumed equal to 13.7
billion years (Spergel et al.\ 2003). Here we note some properties of the
galaxies that we observed, in particular, their $HII$ pattern.

{\em UGC 5427}. This dwarf galaxy with asymmetric $HII$ filaments is
situated at the fartherst western outskirts of the CVnI cloud. Its
distance, 7.1 Mpc, was estimated from the luminosity of the brightest
stars.

{\em NGC 3344}. This is one of the rare galaxies in the cloud with a
quite regular spiral pattern. Its distance of 6.9 Mpc determined from the
radial velocity at a Hubble  constant $H_0=72$ km s$^{-1}$Mpc$^{-1}$ may
be considerably underestimated since the galaxy is located in the Local
Velocity Anomaly zone (Tully et al. 1992).

{\em KK 109}. This dwarf system of low surface brightness belongs to an
intermediate type between irregular and spheroidal galaxies like the known
member of the Local Group, LGS-3. Its distance, 4.51 Mpc, was determined
from the tip of the red giant branch, TRGB (Karachentsev et al. 2003).
There are no massive neighbors ($TI=-0.6$) in the vast vicinities of
KK 109, which could cause pushing of gas from this galaxy by the ram pressure.
The $H{\alpha}$-flux shown in Table 2 for the galaxy corresponds only to
the upper limit.

{\em BTS 76 and MGC 6-27-17}. These are two relatively compact bluish galaxies
the shape of which is not notable for great irregularity. They have not
been resolved into stars yet, and the distances to them are estimated from
their radial velocities. In the central parts of both galaxies compact
emission knots are seen.

{\em DDO 113 = KDG 90}.  This is a dwarf system of regular shape and low
surface brightness. It has been resolved into stars with the Hubble Space
Telescope (HST). Judging by
its CM-diagram, DDO 113 contains an old stellar population and can be
classified as dSph. We have not found in it either $HII$ regions or diffuse
$H{\alpha}$ emission. The measurement of its $HI$-flux is hampered by close
neighborhood ($\sim10\arcmin)$ with the bright Im-galaxy NGC 4214, a companion
of which DDO 113 evidently is.

{\em MGC 9-20-131=CGCG 269-049}. In the central part of this dIrr galaxy a
bright $HII$ region is seen. Together with UGC 7298 it is likely to
constitute a physical pair of dwarf galaxies investigated in the $HI$ line
with the Giant Metrewave Radio Telescope (GMRT) (Begum et al. 2006).

{\em UGC 7298  and UGC 7356}. These are irregular galaxies with a low content
of $HI$. Their distances, 4.21 Mpc (UGC 7298) and 7.19 Mpc (UGC 7356),
have recently been measured by Karachentsev et al. (2003) and
Tully et al. (2007). Some faint emission regions are seen in both galaxies.
Possibly, the high $HI$-flux from UGC~7356 is due to contamination from the
bright neighboring spiral NGC~4258 since the anomalously high ratio
$(M_{HI}/L_B = 13.5M_{\odot}/L_{\odot})$ looks inconsistent with the
morphology of this galaxy.

{\em IC 3308=UGC 7505}. From the Tully-Fisher relation with the $HI$ line
width $W_{50}$ = 128 km s$^{-1}$ we derived the galaxy distance to be
12.8 Mpc.

{\em KK 144}. The measurement of $H{\alpha}$-flux is impeded because
of a bright star projected near the galaxy center.

{\em NGC 4395}. This is a Seyfert 1 type galaxy having a star-like nucleus,
which is lost among bright emission regions scattered over the disk. The integrated
$H{\alpha}$-flux in NGC~4395 as well as in the galaxies NGC~4449  and
NGC~4631 was corrected for the incomplete field of view under the
assumption that the $H{\alpha}$-emission in these galaxies is distributed
in such a way as the blue luminosity.

{\em UGCA 281=Mkn 209}. On the western side of this blue compact dwarf
galaxy (BCD), there is a very bright emission knot with a short arc.
Judging by the distance, 5.43 Mpc, measured by Tully et al. (2007) with
HST, this compact starburst galaxy has no close neighbors
and is quite appropriate to be called ``intergalactic $HII$ region''
given by Sargent \& Searle (1970). We note some inconsistency in
the integrated blue magnitudes for this galaxy. So, Papaderos et al. (1996)
obtained $B_T = 14.84$, while Makarova et al. (1997) and Gil de
Paz et al. (2003) give for it 15.14 and 14.15 mag, respectively.
We observed UGCA 281 with the 6-m telescope in the B,V,R bands and obtained
$B_T = 14\fm87\pm0\fm05$. This magnitude measured by M.E.Sharina is finally
presented in Table 2.

{\em DDO 125=UGC 7577}. The diffuse emission regions in this dIrr galaxy form a
knotty filament extending as far as the northern side of the galaxy.

{\em  UGC 7584}. Together with diffuse emission regions, zones of
noticable internal absorption are seen in the galaxy, which is not typical
of dwarf galaxies.

{\em KKH 80}. This is a galaxy of low surface  brightness and of quit
regular shape. In Table 2 only the upper limit of its $H{\alpha}$-flux
is presented. Most likely, KKH~80 belongs to the transition dIrr/dSph
type.

{\em NGC 4449}. The bright galaxy of Magellanic type reveals numerous
powerful starburst sites scattered over the whole disk. Its emission
filament structure has been described in more detail by Hunter \&
Gallagher (1992).

{\em NGC 4460}. This is a lenticular galaxy without signs of spiral
structure. Surprisingly, we find a compact emission disk in its core,
from which diffuse emission protuberances originated along the minor axis.
Using the surface brighness fluctuations, Tonry et al. (2001) derived
its distance to be 9.59 Mpc.

{\em NGC 4627 and NGC 4631}. This is a close in projection pair of dE and Sd
galaxies with distances 9.38 Mpc and 7.66 Mpc, respectively. The elliptical
component shows weak peripheric distortion, which is likely to give
grounds to include this pair into the catalog of peculiar systems (Arp 1966).
However, with errors in the estimation of the distances to the galaxies of
$\sim15$\% they may constitute a physical pair of the M32+M31 type, in which
the dE component underwent "de-gasetion" because of the
influence of a close massive neighbor.

{\em KK 160}. This is a galaxy of the transition dIrr/dSph type with weak
emission in $HI$, but without visible emission in the $H{\alpha}$ line.

{\em DDO 147=UGC 7946}.  This dIrr galaxy has several emission
regions of different degree of compactness. Its distance, 9.9 Mpc, is
estimated from the brightest stars and needs refinement.

{\em KK 166}. This is a dSph galaxy of very low surface brightness, for
which only upper limits of the fluxes in $HI$ and $H{\alpha}$ are available.
This is one of the faintest ($M_B=-10.82$) known members of the CVnI cloud.

{\em UGC 7990}. The distance to this galaxy (20.9 Mpc) derived from the
Tully-Fisher relation is largely different from the estimate made from
its radial velocity (6.9 Mpc). The galaxy has not been resolved into
stars yet.

{\em UGC 8215}.  To the east of this dIrr galaxy there is a nearly star-like
emission knot.

{\em NGC 5023}. This isolated late-type spiral galaxy seen almost edge-on
was observed with ACS at HST by Seth et al. (2005), who estimated its
distance from the TRGB luminosity to be 6.61 Mpc. All visible $H{\alpha}$
emission from NGC 5023  is concentrated in its disk without signs of
extra-planar gas motions.

{\em NGC 5195}. This is an elliptical component of the famous interacting
pair M51. Its diffuse $H{\alpha}$ emission is distributed over
the galaxy body  non-uniformly.  A considerable part of the
$H{\alpha}$-flux is concentrated in the circumnuclear region.
Some uncertainty in the evaluation of the integrated flux from NGC 5195
is introduced by the $HII$ regions of the spiral arm of NGC 5194  crossing the
elliptical component of the pair.

{\em  NGC 5229}. This is a late-type spiral galaxy seen almost edge-on. The
edges of its disk are slightly curved to the opposite sides rendering the
galaxy an ``integral-like'' shape.

{\em UGC 8638}. This is a compact galaxy whose main $H{\alpha}$-emission
comes from several compact $HII$ regions near its center.

{\em DDO 181=UGC 8651}.  The galaxy has a curved bow-like shape similar to
DDO 165. The basic $H{\alpha}$-flux in it is emitted from a very bright
$HII$ region at the eastern edge.

{\em Holmberg IV=UGC 8837}. This is an irregular galaxy whose
$H{\alpha}$-emission is concentrated in compact $HII$ regions without signs
of a diffuse component. Judging by the distance of 6.83 Mpc measured
by Tully et al. (2007), it is a satellite of the bright spiral galaxy M101.

{\em UGC 8882}. According to Rekola et al. (2005), this is a dE
galaxy at a distance of 8.3 Mpc with a compact nucleus and without
visible emission in the lines $HI$ and $H{\alpha}$.

{\em KK 230}. This is an isolated dIrr galaxy of low surface brightness
situated between the  CVnI cloud and the Local Group at a distance of
1.92 Mpc. No signs of $H{\alpha}$-emission are seen.
The distribution of neutral hydrogen in it with high angular resolution
has been investigated by Begum et al. (2006).

{\em KKH 87}.  This dIrr galaxy is a likely companion to M101. Almost all
$H{\alpha}$-flux of it comes from compact $HII$ regions without signs of
a diffuse component.

{\em DDO 190=UGC 9240}. The periphery of the galaxy has quite a regular shape.
Compact and diffuse emission knots occupy the central and southern
areas of the galaxy. Being an isolated dwarf system, DDO~190 is an
expressive example of a star formation burst not triggered by external
tides.

{\em KKR 25}. This is an isolated dSph galaxy of very low surface
brightness, near which a bright star is projected. KKR~25 is located between
the CVnI cloud and the Local Group at a distance of 1.86 Mpc. Its radial
velocity of +68 km c$^{-1}$ measured by Huchtmeier et al. (2003), has not
been corroborated by deeper observations with GMRT (Begum \& Chengalur, 2005)
who estimated only the upper limit of the $HI$ flux. On the northern side
of KKR~25 one can see in $H{\alpha}$ a faint knot, the nature of which
can be established by spectral observations.

\section{External $SFR$ comparison}
In order to test our photometry, we compared the data on $SFR$ from Table 2
against previous flux measurements. We have found in the literature
eleven cases where $H{\alpha}$-fluxes in the galaxies of the  CVnI cloud
were measured by other authors. The summary of these  data is presented
in Table 3 where the index ``6m'' at log[$SFR$] corresponds to our
measurements, while the index ``oth'' corresponds to other sources indicated
in the last column. The literature $H{\alpha}$-fluxes were corrected
for the external (Galactic) and internal absorption in the manner
described above and reduced to the galaxy distances given in Table 2.
The mean offset is $\Delta$log[SFR]=$-0.03\pm0.01$ in the sense that our
study finds about 7\% lower fluxes than literature studies, possibly
due to the lower effective apertures used. The eleven points have
an RMS scatter of 0.04 dex about the regression line. Given the above
mentioned uncertainties and differences in reduction procedures,
the agreement found in Table 3 seems to be generally good.

\section{Discussion}
As it has been noted above, the population of the CVnI cloud is distinguished
among other nearby, but virialized groups by two basic features: a) most
galaxies in the cloud ($\sim70)$\% do not actually interact with their
closest neighbors, having negative tidal indices; b) the majority of the
cloud objects ($\sim70)$\% are galaxies of low luminosity classified as
$T=10$ (=Irr) or 9 (=Im, BCD). Evidently, both these properties are
interconnected being due to the slow rate of the dynamical evolution of
the CVnI volume under its low density contrast. At the present time, we
have at our disposal a complete set of data on star formation rates and
gas reserves in all known members of the cloud. Their $H{\alpha}$ images
presented in Fig.1 exhibit that galaxies of one and the same morphological
type possess an enormous variety of emission patterns, being almost
all under the conditions of dynamical isolation ($TI<0)$. This
property can be easely understood when the star formation process in
galaxies, in particular its rate, is mainly governed by the internal
but not the external mechanisms (tidal triggering).

The distribution of $SFR$ for all 78 galaxies in the CVnI cloud as a function
of their blue absolute magnitude is presented in Fig.2 by filled circles.
The galaxies with only the upper limit of the $H{\alpha}$-flux
are given by open circles. For comparison, we draw here similar data on
the complete set of 41 galaxies situated
in the nearby virialized group around M81 (Karachentsev
\& Kaisin, 2007). They are shown in this diagram by filled and open
squares. As can be seen, the bright members of the cloud and the group
follow a common linear relationship $SFR\propto L_B$ displayed by the line.
The differences between the samples are not large both from the integrated
$SFR$ range and from the scatter of specific $SFR$ per unit luminosity.
Faint dwarfs in the two samples with $M_B > -13^m$ demonstrate a systematic
offset from the main sequence. However, their displacement in $SFR$ can
be essentially reduced in a new scenario of stellar evolution proposed
by Weidner \& Kroupa, 2005.

Another important diagram (Fig.3) illustrates a relationship between $SFR$
and total hydrogen mass $M_{HI}$ of galaxies. The members of the CVnI cloud
and those of M81 group are shown here by the same symbols as in Fig.2.
As have been already noted by many authors: Kennicutt (1989, 1998),
Taylor \& Webster (2005), Tutukov (2006), spiral and irregular galaxies
show a steeper dependence of $SFR$ on $M_{HI}$ than on luminosity $L_B$,
namely $SFR\propto M_{HI}^{1.5}$. This testifies that dIrr galaxies
preserve a relatively larger amount of gas than spirals to maintain
star formation with the now observed rates. This diagram also
shows that the members of the CVnI cloud and M81 group are mutually
well mixed as in the previous plot.

To estimate the evolutionary status of normal and dwarf galaxies in the CVnI
cloud, we used the data on dimensionless parameters in the last two columns
of Table 2: $p_*=\log[(SFR)T_0/L_B]$ and $f_*=\log[M_{HI}/(SFR)T_0]$,
which characterize the past and future of the star formation process
in a galaxy on the assumption of permanent star formation rate. The
evolutional ``past-future'' diagram for the observed galaxies is presented
in Fig.4. The members of the CVnI cloud are shown by circles, and
the members of the group M81 are marked by squares. The open symbols
correspond to the cases where observations give only the upper limit
of the flux in $H{\alpha}$ or $HI$.

 The distribution of  galaxies in the diagnostic \{$p_*,f_*$\} diagram
shows some interesting properties. The members of the CVnI cloud, on
the whole, are located symmetrically enough with respect to the origin of
coordinates, having median values $p_*=+0.02$ and $f_*=-0.03.$ This
means that the observed star formation rates in the CVnI galaxies proved
to be quite sufficient to reproduce their observed luminosity (baryon
mass). Moreover, the cloud galaxies possess gas reserves sufficient
to maintain the observed star formation rates during one more Hubble
time $T_0$, being just in the middle of their evolutionary path.
For comparison, the median values $p_*$ and $f_*$ for the M81 group members
are $-0.30$ and $+0.12$, respectively, i.e. their typical value of $SFR$
per unit luminosity is twice as low as in the galaxies of CVnI.
But this difference is easily explicable by the presence around M81 of
a lot of dSphs in which the current star formation can be suppressed
by tidal stripping. The CVnI cloud contains only 3 or 4 such "extinct"
dwarfs: KK109, DDO113, KK166 and, probably, UGC8882.

In contrast to spiral galaxies, dIrr galaxies of low luminosity exhibit
multiple episodes of global star formation (Dohm-Palmer et al. 2002,
Dolphin et al. 2003, Skillman 2005, McConnachie et al. 2005, Young
et al. 2007). Stinson et al. 2007 simulated the collaps of isolated
dwarf galaxies with the effects of supernova feedback and showed that
star formation in them occurs in the form of bursts rather than in the
mode of a sluggish process. The observed $H{\alpha}$ flux in galaxies
represents its current $SFR$ only over the past $\sim10$ Myr (Bell \&
Kennicutt, 2001, Annibali et al. 2007). The difference between the observed
"momentary" and the secular value of $SFR$ averaged over $T_0$ will lead
to scatter of flashing and dimming dIrr galaxies in the diagram
\{$p_*,f_*$\} along the diagonal line $p_*=-f_*$.
This tendency is actually seen in Fig.4 both for the CVnI cloud and for
the M81 group members. These data permit us to estimate that the global
star formation rate in the dIrrs can be varied with time  by about
an order of magnitude. The most expressive representatives of a dwarf
galaxy at a burst stage are UGCA~281=Mkn209 and UGC~6541=Mkn178,
in which the specific $SFR$ per unit luminosity is 16 times and 7 times
higher than the average value. The known ``exploding'' galaxy M82
in the M81 group has a bit lower specific $SFR$.

Among 500 galaxies of the Local volume there are only 6 galaxies with
the hydrogen mass-to-luminosity ratio greater 5 $M_{\sun}/L_{\sun}$.
Surprisingly, half of these  semi-gaseous galaxies: DDO~154, NGC~3741
and UGCA~292 reside in the CVnI cloud. All three of them are located
in the upper right quadrant of Fig.4. Considering the M81 group,
Karachentsev \& Kaisin (2007) have found in this quadrant only one
peculiar object, the dark $HI$ cloud HIJASS with a rather
uncertain estimate of $SFR$. Probably, galaxies start their evolution
just from this quadrant, [$p_*>0,f_*>0]$,
converting their initial gaseous mass into stars.

There is a widespread point of view that close galaxy encounters trigger
enhanced star formation in them. However, Hunter \& Elmegreen, 2004
and Noeske et al. 2001 find no correlation between star formation
activity in galaxy and its proximity to other neighboring galaxies.
Telles \& Maddox, 2000 and James et al. 2004 found that bursting
dwarf galaxies inhabit slightly lower density environments than those of
denser field. Our data are in agreement with such a conclusion. The
left-hand diagram of Fig.5 presents the distribution of members of the CVnI
cloud (circles) and of the M81 group (squares) versus their specific $SFR$ and
tidal index. Here the solid regression line has a slope of ($-0.01\pm0.03$)
corresponding to the galaxies detected in the $H\alpha$ line, while the dashed
regression line with a slope of ($-0.07\pm0.05$) corresponds to the whole
sample of $78+41$ galaxies including those with upper limits of their
$H\alpha$ flux.
This diagram does not show significant difference in $SFR$ for
galaxies in groups ( $TI > 0$) as compared to isolated ones.
In particular, the strongly disturbed system Garland near NGC~3077 and
the isolated blue galaxy UGCA~281 have almost the same
extremely high values of $p_*$.

The right diagram of Fig.5 presents the time of exhaustion of the
available reserves of $HI$ gas, $f_*$, for the CVnI cloud and the M81 group
galaxies versus their tidal index. The solid and dashed regression lines with
slopes ($-0.10\pm0.04$) and ($-0.01\pm0.05$) correspond to the galaxies
detected in the $H\alpha$  and $HI$ lines, and to the whole sample of galaxies,
respectively.
Despite the considerable dispersion,
a trend of diminishing of the value of $f_*$ from isolated galaxies
towards interacting ones is seen, which reflects the known tendency
of rising $HI$-deficiency with increasing density of its environment
(Giovanelli \& Haynes, 1991).

We regard the isolated ($TI=-0.7$)
lenticular galaxy NGC~4460 to be the most intriguing object in the
considered sample. A powerful star formation burst in
its center exhasts all available reserve of gas just for 170 Myr.
Possibly, we observe here a rare event of interaction of the S0 galaxy
with an intergalactic $HI$ cloud. Such dark completely starless clouds have
already been detected in the nearby groups M81 (Boyce et al. 2001), Leo-I
(Schneider 1985) and the nearby Virgo cluster (Minchin et al. 2005).

\section{Conclusions}
A systematic survey of $H{\alpha}$-emission in the nearest scattered cloud
CVnI shows that in most of its galaxies the process of active star
formation is on despite the low density contrast of this cloud and
rather rare interaction between its members. By making full use of our
$H{\alpha}$-survey, we can estimate the mean density of $SFR$ in the
cloud. A conical volume in the distance range from 2 to 10 Mpc,
resting in a sky region of $\sim1500$ square degrees, makes 153 Mpc$^3$.
In this volume we have a summary value $\Sigma(SFR)=18.6M_{\sun}$yr$^{-1}$,
which yields an average density of star formation rate
$\dot{\rho}_{SFR}=0.12M_{\sun}$yr$^{-1}$Mpc$^{-3}$.
The obtained value turns out
to be a little less than $\dot{\rho}_{SFR}=0.165M_{\sun}$yr$^{-1}$Mpc$^{-3}$
for a ``sell of homogeneity'' embracing the group M81
(Karachentsev \& Kaisin, 2007). According to Nakamura et al.
(2004), Martin et al. (2005) and Hanish et al. (2006), the average star
formation rate per 1 Mpc$^3$ at the present epoch (z=0) is
(0.02--0.03)$M_{\sun}$yr$^{-1}$Mpc$^{-3}$. Consequently, the CVnI cloud
has excess of $SFR$ density  4--6 times higher in comparison with
the mean global quantity, which roughly corresponds to the cloud
density contrast on the sky, $\Delta N/N\sim4$.
Herefrom we conclude that the CVnI cloud is characterised by a usual norm
of star formation in its galaxies.

\acknowledgements{
We would like to thank Margarita Sharina for her help with the photometry
of UGCA 281. We are also grateful to B.\ Tully for useful discussions.
Support associated with HST program 10905 was provided by
NASA through a grant from the Space Telescope Science Institute,
which is operated by the Association of Universities for Research in
Astronomy, Inc., under NASA contract NAS5--26555. This work was also
supported by RFFI grant 07--02--00005 and grant DFG-RFBR 06--02-04017.}

{}
\input{t1.tex}
\input{t2.tex}
\input{t3.tex}

\setcounter{figure}{1}
\begin{figure}[p]
\centering
\includegraphics[width=10cm]{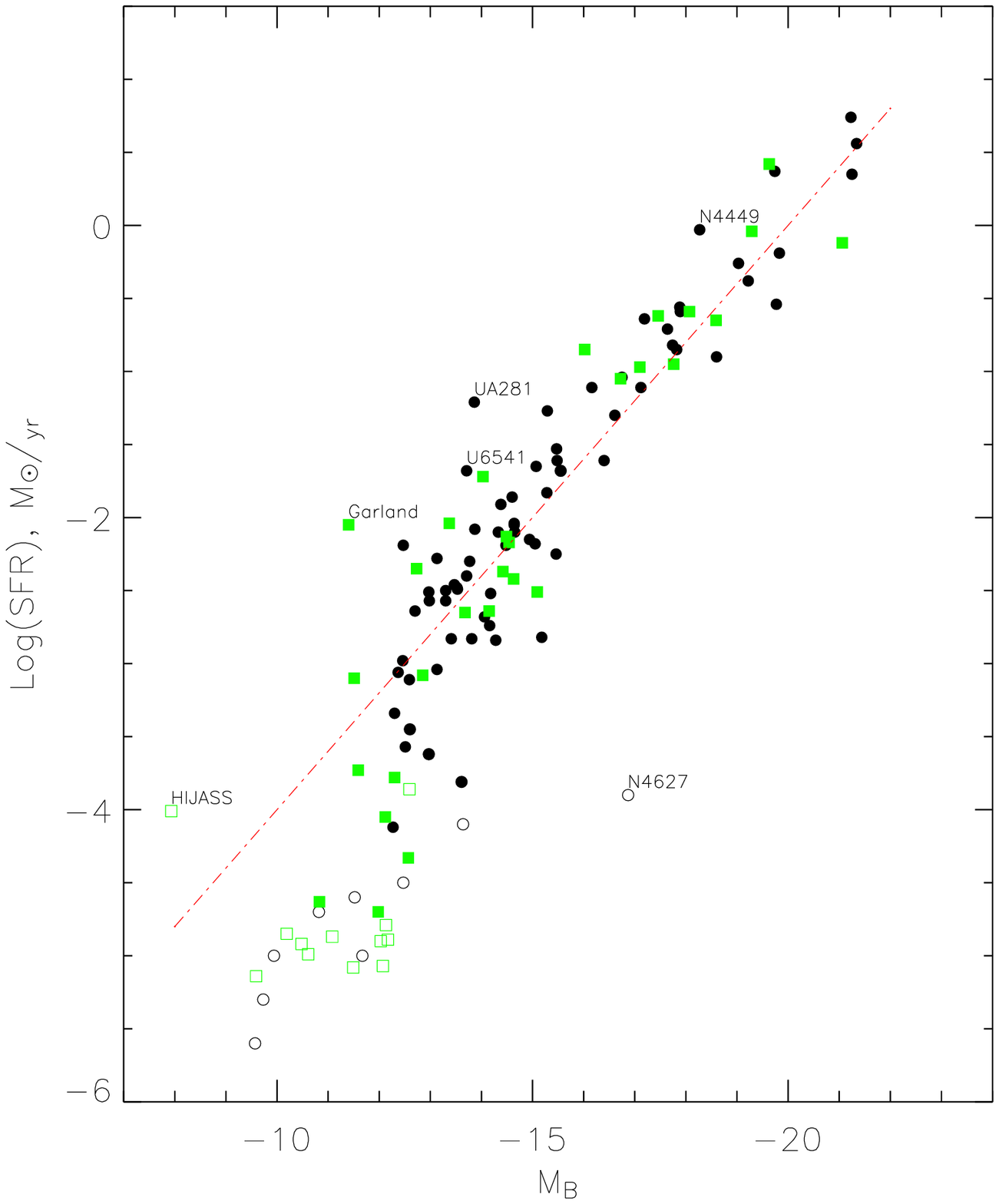}
\caption{Star formation rate versus blue absolute magnitude for 78
galaxies in the Canes Venatici I cloud (circles) and 41 members of the M81
group (squares). The open symbols indicate the galaxies with only upper
limit of their $SFR$. The line corresponds to a constant $SFR$ per unit
luminosity.}
\end{figure}

\setcounter{figure}{2}
\begin{figure}[p]
\centering
\includegraphics[width=10cm]{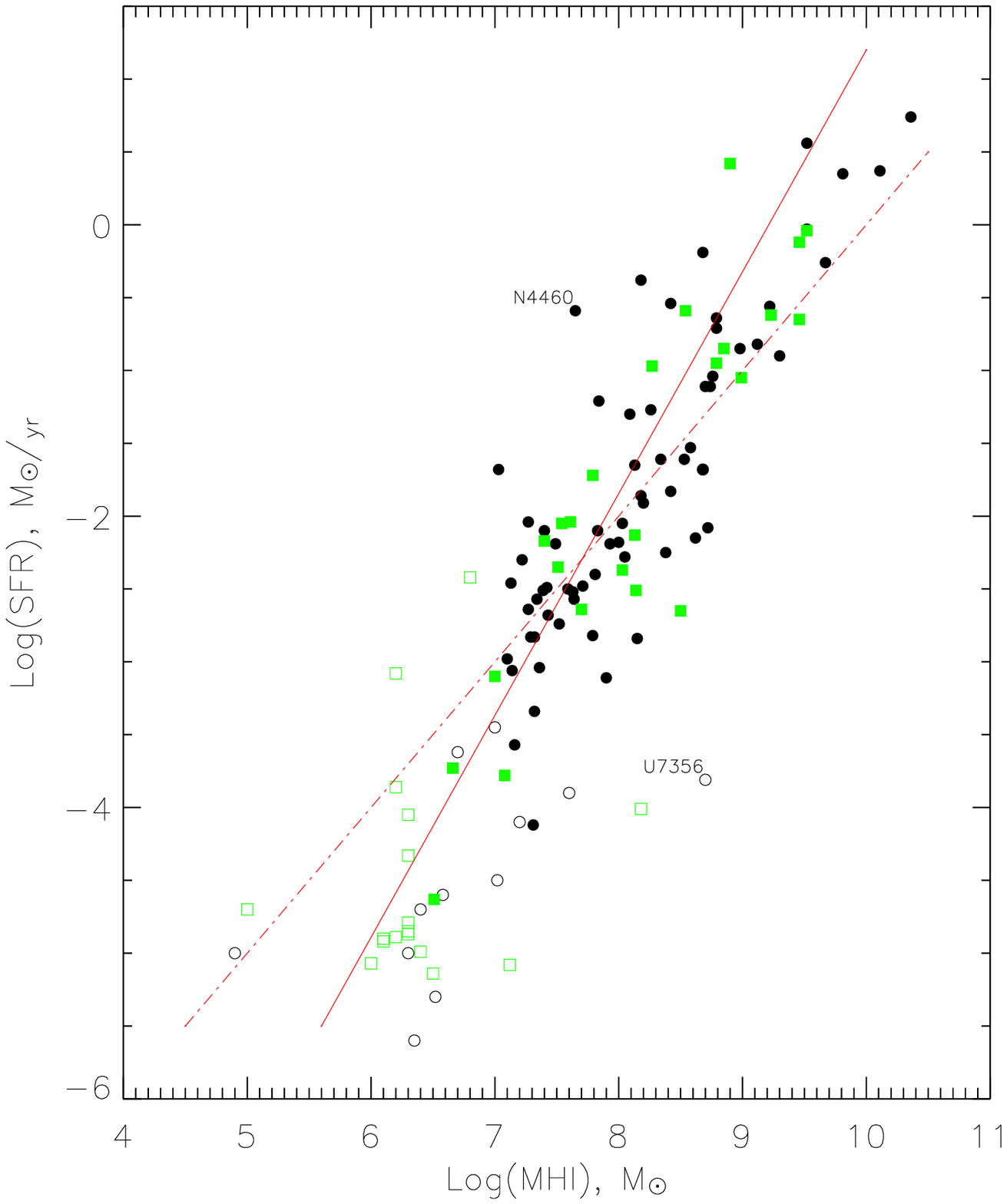}
\caption{Star formation rate versus neutral hydrogen mass for galaxies
in the CVnI cloud (circles) and the M81 group (squares). The galaxies
with upper limit of $SFR$ or $M_{HI}$ are indicated by open symbols. The
dashed line corresponds to a fixed $SFR$ per unit hydrogen mass and the
solid line traces the relationship $SFR\propto M_{HI}^{1.5}$.}
\end{figure}

\setcounter{figure}{3}
\begin{figure}[p]
\includegraphics[width=20cm]{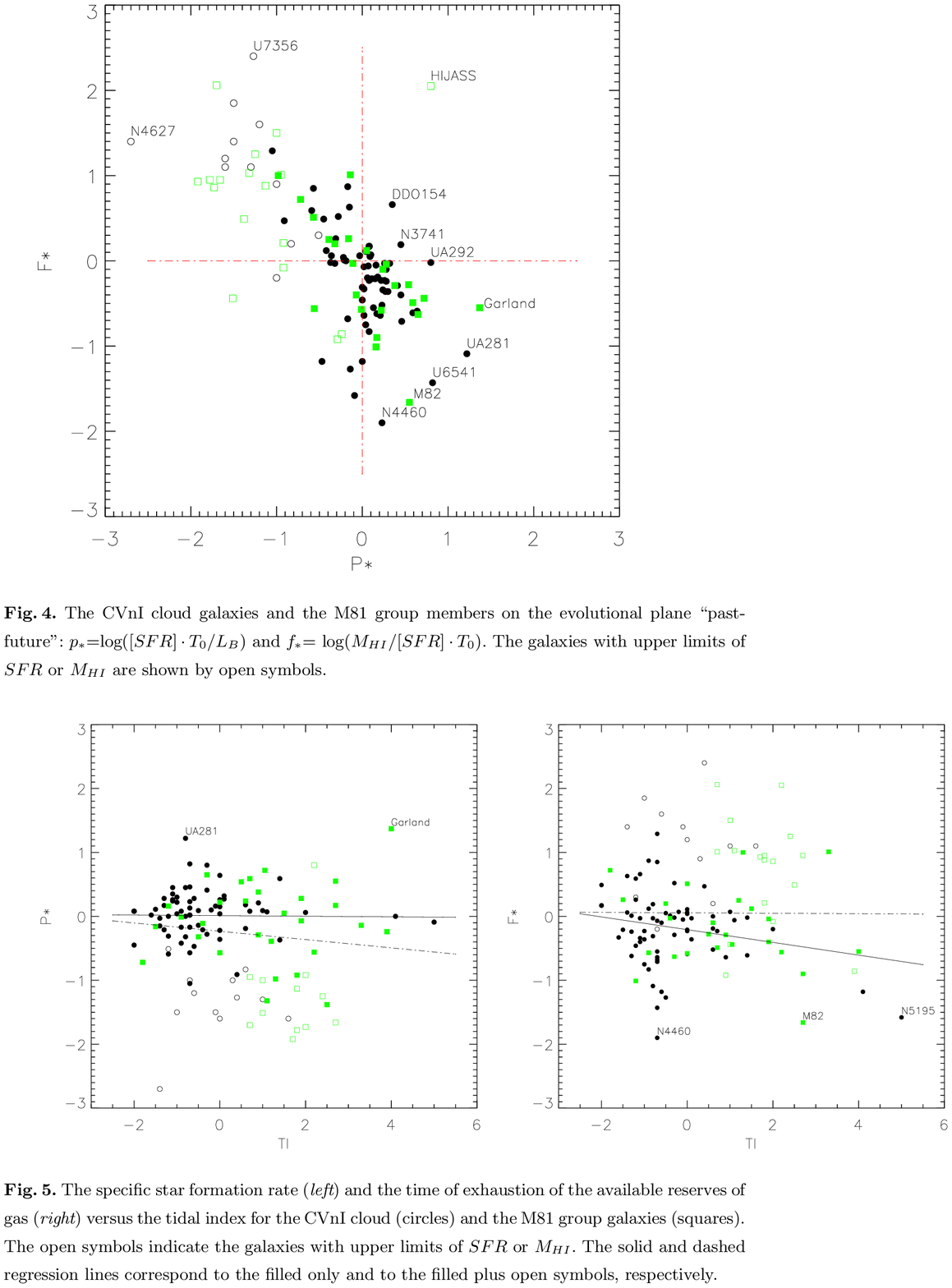}
\end{figure}
\end{document}

%% file: t1.tex
\clearpage
\footnotesize
\onecolumn
\begin{table}
\caption{The observational Log}
\begin{tabular}{lcc} \\ \hline
Galaxy      &    Date      & $T_{exp}$ \\
\hline
U5427       &  03/02/2005  & 1200      \\
N3344       &  04/02/2005  &\ 300      \\
KK109       &  05/02/2005  & 1200      \\
BTS76       &  04/02/2005  & 1200      \\
MCG6-27-17  &  05/02/2005  & 1200      \\
DDO113      &  27/01/2004  & 1200      \\
MCG9-20-131 &  06/02/2005  & 1200      \\
U7298       &  06/02/2005  & \ 900      \\
U7356       &  05/02/2005  & 1200      \\
I3308       &  04/02/2005  & \ 600      \\
KK144       &  04/02/2005  & 1200      \\
N4395       &  06/02/2005  & \ 300      \\
UA281       &  24/05/2006  & \ \ 300:     \\
DDO125      &  18/05/2005  & 1200      \\
U7584       &  04/02/2005  & 1200      \\
KKH80       &  04/02/2005  & 1200      \\
N4449       &  24/05/2006  &  \ 500      \\
DDO127      &  05/02/2005  & 1200      \\
U7605       &  05/02/2005  & 1200      \\
N4460       &  06/02/2005  &  \ 600      \\
KK149       &  06/02/2005  & 1200      \\
U7639       &  06/02/2005  & 1200      \\
KK151       &  06/02/2005  & 1200      \\
N4627       &  06/02/2005  & \ 600      \\
N4631       &  22/05/2006  &  \ \ 900:     \\
KK160       &  22/03/2006  & \ 1200:     \\
DDO147      &  05/02/2005  & 1200      \\
KK166       &  22/03/2006  & \ 1200:     \\
U7990       &  05/02/2005  & 1200      \\
U8215       &  16/05/2005  & 1200      \\
N5023       &  04/02/2005  & \ 600      \\
N5195       &  05/05/2005  & \ 500      \\
U8508       &  17/03/2001  & 1200      \\
N5229       &  06/02/2005  & 1200      \\
U8638       &  06/02/2005  & 1200      \\
DDO181      &  24/05/2006  & \ 500      \\
Holmberg IV &  29/01/2004  & \ 600      \\
U8882       &  21/03/2006  &  \ 1200:     \\
KK230       &  03/02/2005  & 1200      \\
KKH87       &  29/01/2004  & 1200      \\
DDO190      &  24/05/2006  & \ 400      \\
KKR25       &  25/03/2001  &   \ \ 600:     \\
\hline
\end{tabular}
\end{table}

%% file: t2.tex
\clearpage
\footnotesize
\onecolumn
\tablecaption{Basic parameters of galaxies in the Canes Venatici I cloud.}
\tablehead{\hline
\multicolumn{1}{c}{Name}   &
\multicolumn{1}{c}{RA (J2000) Dec} &
\multicolumn{1}{c}{T}&
\multicolumn{1}{c}{$TI$} &
\multicolumn{1}{c}{$V_{LG}$}&
\multicolumn{1}{c}{$D$}&
\multicolumn{1}{c}{$M_B$} &
\multicolumn{1}{c}{log$M_{HI}$}&
\multicolumn{1}{c}{log$F$}   &
\multicolumn{1}{c}{log$SFR$} &
\multicolumn{1}{c}{$p_*$}  &
\multicolumn{1}{c}{$f_*$}    \\
  &  &  &  & &
\multicolumn{1}{c}{Mpc} &
\multicolumn{1}{c}{mag} &
\multicolumn{1}{c}{$M_{\sun}$} &
 &
\multicolumn{1}{c}{$M_{\sun}$ yr$^{-1}$} &      &       \\
\hline
\multicolumn{1}{c}{(1)}&
\multicolumn{1}{c}{(2,3)}&
\multicolumn{1}{c}{(4)}&
\multicolumn{1}{c}{(5)}&
\multicolumn{1}{c}{(6)}&
\multicolumn{1}{c}{(7)}&
\multicolumn{1}{c}{(8)}&
\multicolumn{1}{c}{(9)}&
\multicolumn{1}{c}{(10)}&
\multicolumn{1}{c}{(11)}&
\multicolumn{1}{c}{(12)}&
\multicolumn{1}{c}{(13)} \\ \hline
}
\par
\begin{supertabular}{llrrrlrrcrrr}
U5427  &100441.0+292159 & 8& $-$1.2& 424 & 7.1  &$-$14.48&  7.49& $-$13.03& $-$2.19& $ $0.00& $-$0.46 \\
U5672  &102820.9+223417 &10& $-$0.7& 428 & 6.3  &$-$14.65&  7.40& $ $  a  & $-$2.10& $ $0.02& $-$0.64 \\
N3274  &103217.1+274007 & 6& $-$0.3& 461 & 6.5  &$-$16.16&  8.74& $ $  a  & $-$1.11& $ $0.41& $-$0.29 \\
N3344  &104330.2+245525 & 4& $-$1.5& 498 & 6.9  &$-$19.03&  9.67& $-$11.09& $-$0.26& $ $0.11& $-$0.21 \\
U6541  &113329.1+491417 &10& $-$0.7& 304 & 3.89 &$-$13.71&  7.03& $ $  b  & $-$1.68& $ $0.82& $-$1.43 \\
N3738  &113548.6+543122 &10& $-$1.0& 305 & 4.90 &$-$16.61&  8.09& $ $  a  & $-$1.30& $ $0.04& $-$0.75 \\
N3741  &113606.4+451707 &10& $-$0.8& 264 & 3.03 &$-$13.13&  8.05& $ $  a  & $-$2.28& $ $0.45& $ $0.19 \\
KK109  &114711.2+434019&$-$1& $-$0.6& 241 & 4.51 &$-$9.73&  6.52& $-$15.7:& $-$5.3:& $-$1.2:& $ $1.6: \\
DDO99  &115053.0+385250 &10& $-$0.5& 248 & 2.64 &$-$13.52&  7.71& $ $  a  & $-$2.48& $ $0.09& $ $0.05 \\
BTS76  &115844.1+273506 &10& $-$1.2& 451 & 6.3  &$-$12.60&  7.0:& $-$14.17& $-$3.45& $-$0.51& $ $0.3: \\
N4068  &120402.4+523519 &10& $-$1.0& 290 & 4.31 &$-$15.07&  8.13& $ $  a  & $-$1.65& $ $0.30& $-$0.36 \\
MCG627 &120956.4+362607 &10& $ $0.6& 341 & 4.7  &$-$12.97&  6.7:& $-$14.09& $-$3.62& $-$0.83& $ $0.2: \\
N4144  &120959.3+462726 & 6& $ $0.9& 319 & 7.41 &$-$17.64&  8.79& $ $  a  & $-$0.71& $ $0.21& $-$0.64 \\
N4163  &121208.9+361010 &9& $ $1.4& 164 & 2.96 &$-$13.81&  7.29& $ $  a  & $-$2.83& $-$0.37& $-$0.02 \\
N4190  &121344.6+363760 &10& $ $0.0& 234 & 3.5  &$-$14.33&  7.83& $ $  a  & $-$2.10& $ $0.15& $-$0.21 \\
DDO113 &121457.9+361308&$-$1& $ $1.6& 283 & 2.86 &$-$11.67&  6.3:& $-$15.0:& $-$5.0:& $-$1.6:& $ $1.1: \\
N4214  &121538.9+361939 & 9& $-$0.7& 295 & 2.94 &$-$17.19&  8.79& $ $  c  & $-$0.64& $ $0.46& $-$0.71 \\
MCG920 &121546.7+522315 &10& $-$0.7& 245 & 3.4  &$-$12.46&  7.10& $-$13.17& $-$2.98& $ $0.02& $-$0.07 \\
U7298  &121628.6+521338 &10& $-$0.7& 255 & 4.21 &$-$12.27&  7.31& $-$14.49& $-$4.12& $-$1.05& $ $1.29 \\
N4244  &121729.9+374827 & 6& $-$0.0& 255 & 4.49 &$-$18.60&  9.30& $ $  d  & $-$0.90& $-$0.36& $ $0.06 \\
N4258  &121857.5+471814 & 4& $-$0.7& 507 & 7.83 &$-$21.25&  9.81& $ $  e  & $ $0.35& $-$0.17& $-$0.68 \\
U7356  &121909.1+470523 &10& $ $0.4& 330 & 7.19 &$-$13.61&  8.7:& $-$14.64& $-$3.81& $-$1.27& $ $2.4: \\
I3308  &122517.9+264253 & 7& $-$2.:& 277 &12.8  &$-$15.55&  8.68& $-$13.10& $-$1.68& $ $0.08& $ $0.17 \\
KK144  &122527.9+282857 &10& $-$0.9& 453 & 6.3  &$-$12.59&  7.90& $-$13.83& $-$3.11& $-$0.17& $ $0.87 \\
N4395  &122549.8+333246 & 8& $ $0.1& 315 & 4.67 &$-$17.88&  9.22& $-$11.04& $-$0.56& $ $0.27& $-$0.36 \\
UA281  &122616.0+482931 & 9& $-$0.8& 349 & 5.43 &$-$13.86&  7.84& $-$11.80& $-$1.21& $ $1.22& $-$1.09 \\
DDO126 &122705.1+370833 &10& $ $0.1& 231 & 4.87 &$-$14.38&  8.20& $ $  a  & $-$1.91& $ $0.32& $-$0.03 \\
DDO125 &122741.8+432938 &10& $-$0.9& 240 & 2.54 &$-$14.16&  7.52& $-$12.67& $-$2.74& $-$0.42& $ $0.12 \\
U7584  &122802.9+223522 & 9& $-$0.1& 545 & 7.6  &$-$13.30&  7.59& $-$13.39& $-$2.50& $ $0.16& $-$0.05 \\
KKH80  &122805.4+221727 &10& $-$0.1& 542 & 7.5  &$-$12.47&  7.02& $-$15.4:& $-$4.5:& $-$1.5:& $ $1.4: \\
N4449  &122811.2+440540 & 9& $-$0.0& 249 & 4.21 &$-$18.27&  9.52& $-$10.43& $-$0.03& $ $0.64& $-$0.59 \\
DDO127 &122828.5+371400 & 9& $-$0.7& 291 & 6.9  &$-$14.28&  8.15& $-$13.64& $-$2.84& $-$0.57& $ $0.85 \\
U7605  &122839.0+354305 &10& $ $0.7& 317 & 4.43 &$-$13.53&  7.42& $-$12.90& $-$2.49& $ $0.08& $-$0.23 \\
N4460  &122845.8+445152 & 1& $-$0.7& 542 & 9.59 &$-$17.89&  7.65& $-$11.72& $-$0.59& $ $0.23& $-$1.90 \\
KK149  &122852.3+421040 &10& $-$0.8& 446 & 6.2  &$-$14.06&  7.43& $-$13.39& $-$2.68& $-$0.32& $-$0.03 \\
U7639  &122953.0+473148 &10& $ $0.4& 446 & 7.1  &$-$15.18&  7.79& $-$13.64& $-$2.82& $-$0.91& $ $0.47 \\
KK151  &123023.8+425405 & 9& $-$0.4& 479 & 6.7  &$-$13.41&  7.32& $-$13.60& $-$2.83& $-$0.21& $ $0.01 \\
DDO133 &123253.0+313221 &10& $-$1.1& 321 & 6.1  &$-$15.47&  8.58& $ $  c  & $-$1.53& $ $0.26& $-$0.03 \\
Arp211 &123721.3+384443 &10& $-$0.7& 484 & 6.70 &$-$13.47&  7.13& $ $  b  & $-$2.46& $ $0.13& $-$0.55 \\
UA292  &123840.0+324600 &10& $-$0.3& 306 & 5.0  &$-$12.47&  7.93& $ $  f  & $-$2.19& $ $0.80& $-$0.02 \\
N4627  &124159.7+323425&$-$3& $-$1.4:&541 & 9.38 &$-$16.87&  7.6:& $-$15.0:& $-$3.9:& $-$2.7:& $ $1.4: \\
N4631  &124208.0+323229 & 7& $-$1.1:&605 & 7.66 &$-$19.74& 10.11& $-$11.04& $ $0.37& $ $0.45& $-$0.40 \\
I3687  &124215.1+383007 &10& $ $1.1& 385 & 4.57 &$-$14.64&  8.03& $ $  a  & $-$2.05& $ $0.07& $-$0.06 \\
KK160  &124357.4+433941 &10& $ $1.0& 346 & 4.8  &$-$11.52&  6.58& $-$15.1:& $-$4.6:& $-$1.3:& $ $1.1: \\
DDO147 &124659.8+362835 &10& $-$1.4& 351 & 9.9  &$-$14.94&  8.62& $-$13.26& $-$2.15& $-$0.15& $ $0.63 \\
KK166  &124913.3+353645&$-$3& $ $0.3&     & 4.74 &$-$10.82&  6.4:& $-$15.15:&$-$4.7:& $-$1.0:& $ $0.9: \\
U7990  &125027.0+282107 &10& $-$2.:& 495 &20.9  &$-$15.46&  8.38& $-$14.01& $-$2.25& $-$0.45& $ $0.49 \\
N4736  &125053.5+410710 & 2& $-$0.5& 353 & 4.66 &$-$19.83&  8.68& $ $  e  & $-$0.19& $-$0.14& $-$1.27 \\
DDO154 &125405.2+270855 &10& $-$1.1& 355 & 3.91 &$-$13.87&  8.72& $ $  a  & $-$2.08& $ $0.35& $ $0.66 \\
N4826  &125644.2+214105 & 2& $-$0.6& 364 & 4.44 &$-$19.77&  8.42& $ $  g  & $-$0.54& $-$0.47& $-$1.18 \\
I4182  &130549.3+373621 & 9& $ $0.6& 356 & 4.70 &$-$16.40&  8.53& $ $  a  & $-$1.61& $-$0.19& $ $0.00 \\
U8215  &130803.6+464941 &10& $-$0.3& 297 & 4.55 &$-$12.30&  7.32& $-$13.77& $-$3.34& $-$0.28& $ $0.52 \\
N5023  &131211.9+440219 & 6& $-$1.6& 476 & 6.61 &$-$17.12&  8.70& $-$12.03& $-$1.11& $ $0.02& $-$0.33 \\
DDO167 &131322.8+461911 &10& $ $0.0& 243 & 4.19 &$-$12.70&  7.27& $ $  c  & $-$2.64& $ $0.26& $-$0.23 \\
DDO168 &131428.6+455510 &10& $-$0.0& 273 & 4.33 &$-$15.28&  8.42& $ $  a  & $-$1.83& $ $0.04& $ $0.11 \\
DDO169 &131530.7+472947 &10& $-$0.2& 345 & 4.23 &$-$13.71&  7.81& $ $  a  & $-$2.40& $ $0.10& $ $0.07 \\
N5204  &132936.4+582504 & 9& $-$1.1& 341 & 4.65 &$-$16.75&  8.76& $ $  a  & $-$1.04& $ $0.24& $-$0.34 \\
N5194  &132956.0+471404 & 5& $ $4.1& 555 & 8.0  &$-$21.34&  9.52& $ $  g  & $ $0.56& $ $0.00& $-$1.18 \\
N5195  &132958.7+471605&$-$1& $ $5.0& 558 & 8.02 &$-$19.22&  8.18& $-$11.33& $-$0.38& $-$0.09& $-$1.58 \\
U8508  &133044.4+545436 &10& $-$1.0& 186 & 2.56 &$-$12.98&  7.34& $-$12.51& $-$2.57& $ $0.22& $-$0.23 \\
N5229  &133402.9+475455 & 7& $-$0.6& 460 & 5.1  &$-$14.60&  8.18& $-$12.46& $-$1.86& $ $0.28& $-$0.10 \\
N5238  &133442.7+513650 & 9& $-$0.9& 345 & 4.24 &$-$14.64&  7.27& $ $  a  & $-$2.04& $ $0.08& $-$0.83 \\
U8638  &133919.4+244633 & 9& $-$1.3& 273 & 4.27 &$-$13.77&  7.22& $-$12.68& $-$2.30& $ $0.17& $-$0.62 \\
DDO181 &133953.8+404421 &10& $-$1.3& 272 & 3.02 &$-$12.97&  7.39& $-$12.58& $-$2.51& $ $0.28& $-$0.24 \\
DDO183 &135051.1+380116 &10& $-$1.2& 257 & 3.18 &$-$13.13&  7.36& $ $  a  & $-$3.04& $-$0.31& $ $0.26 \\
HolmIV &135445.1+535417 &10& $ $0.6& 276 & 6.83 &$-$15.48&  8.34& $-$12.40& $-$1.61& $ $0.18& $-$0.19 \\
U8833  &135448.7+355015 &10& $-$1.4& 285 & 3.12 &$-$12.37&  7.14& $ $  e  & $-$3.06& $-$0.03& $ $0.06 \\
U8882  &135714.6+540603&$-$1& $ $0.0&     & 8.3  &$-$13.64&  7.2:& $-$15.04:&$-$4.1:& $-$1.6:& $ $1.2: \\
M101   &140312.8+542102 & 6& $ $0.6& 379 & 7.38 &$-$21.23& 10.36& $ $  e  & $ $0.74& $ $0.23& $-$0.52 \\
N5474  &140502.1+533947 & 8& $ $2.0& 413 & 7.2  &$-$17.74&  9.12& $ $  a  & $-$0.82& $ $0.06& $-$0.20 \\
N5477  &140533.1+542739 & 9& $ $1.4& 443 & 7.7  &$-$15.29&  8.26& $ $  a  & $-$1.27& $ $0.59& $-$0.61 \\
KK230  &140710.7+350337 &10& $-$1.0& 126 & 1.92 &$-$9.57&  6.35& $-$15.32:&$-$5.6:& $-$1.5:& $ $1.85: \\
KKH87  &141509.4+570515 &10& $ $1.0& 473 & 7.4  &$-$13.30&  7.64& $-$13.42& $-$2.57& $ $0.09& $ $0.07 \\
DDO187 &141556.5+230319 &10& $-$1.2& 172 & 2.28 &$-$12.51&  7.16& $ $  a  & $-$3.57& $-$0.59& $ $0.59 \\
N5585  &141948.3+564349 & 4& $-$0.8& 459 & 5.7  &$-$17.82&  8.98& $ $  a  & $-$0.85& $ $0.00& $-$0.31 \\
DDO190 &142443.5+443133 &10& $-$1.3& 263 & 2.79 &$-$14.18&  7.63& $-$12.53& $-$2.52& $-$0.21& $ $0.01 \\
DDO194 &143524.6+571524 &10& $-$0.0& 385 & 8.0  &$-$15.05&  8.00& $ $  a  & $-$2.18& $-$0.22& $ $0.04 \\
KKR25  &161347.6+542216&$-$1& $-$0.7&  68:& 1.86 &$-$9.94&  4.9:& $-$14.64:&$-$5.0:& $-$1.0:& $-$0.2: \\
\hline
\multicolumn{1}{l}{\bf NOTES} & & & & & & &  & & & &            \\
\multicolumn{2}{l}{a $-$  James et al., 2004,}& & &  & &  & & & & & \\
 \multicolumn{2}{l}{b $-$ Gil de Paz et al., 2003,}  & & & & &  & & & & & \\
 \multicolumn{2}{l}{c $-$ Hunter \& Elmegreen, 2004,} & & & & &  & & & & & \\
\multicolumn{2}{l}{d $-$ Strickland et al., 2004,}& &  & & &  & & & & &    \\
\multicolumn{2}{l}{e $-$ Kennicutt et al., 1989} & &  & & &  & & & & &     \\
\multicolumn{2}{l}{f $-$ van Zee, 2000}   & & & & &   & & & & &            \\
\multicolumn{2}{l}{g $-$ Young et al., 1996. } & &  & & &  & & & & &       \\
\hline\hline
\end{supertabular}
\protect

%% file: t3.tex
\clearpage
\setcounter{table}{2}
\begin{table}
\caption{Comparison of SFR estimates for the CVnI galaxies}
\begin{tabular}{llccl} \\ \hline
 Galaxy     &log (SFR)$_{6m}$  & log (SFR)$_{oth}$  &  $\Delta$  &       Source          \\
\hline
NGC4395     &  $-$0.56       &    $-$0.59       &  \ \  0.03  &   Kennicutt et al. 1989        \\
UGCA281     &  $-$1.21       &    $-$1.19       &  $-$0.02  &   Gil de Paz et al. 2003     \\
DDO125      &  $-$2.74       &    $-$2.72       &  $-$0.02  &   Hunter \& Elmegreen 2004   \\
NGC4449     &  $-$0.03       &    $ $\ \ 0.01       &  $-$0.04  &   Kennicutt et al. 1989      \\
NGC4631     &  $ $\ \ 0.37:      &    $ $\ \ 0.42       &  $-$0.05  &   Hoopes et. 1999            \\
NGC4631     &  $ $\ \ 0.37:      &    $ $\ \ 0.41       &  $-$0.04  &   Kennicutt et al. 1989      \\
UGC8508     &  $-$2.57       &    $-$2.51       &  $-$0.06  &   James et al. 2004          \\
DDO181      &  $-$2.51       &    $-$2.47       &  $-$0.04  &   van Zee 2000               \\
HolmbergIV  &  $-$1.61       &    $-$1.57       &  $-$0.04  &   James et al. 2004          \\
DDO190      &  $-$2.52       &    $-$2.43       &  $-$0.09  &   James et al. 2004          \\
DDO190      &  $-$2.52       &    $-$2.56       &  $ $\ \ 0.04  &   van Zee 2000               \\
\hline
\end{tabular}
\end{table}                                   

%% file: text_new.bbl
\begin{thebibliography}{}
\bibitem{}Afanasiev V.L., Gazhur E.B., Zhelenkov S.R. \& Moiseev A.V. 2005,
    Bull.SAO, 58, 90
\bibitem{}Annibali F., Aloisi A., Mack J. et al. 2007, astro-ph/0708.0852
\bibitem{}Arp H. 1966, Atlas of Peculiar Galaxies, ApJS, 14, 1
\bibitem{}Begum A. \& Chengalur J.N. 2005, MNRAS, 362, 609
\bibitem{}Begum A., Chengalur J.N., Karachentsev I.D., Kaisin S.S. \& Sharina M.E.
	  2006, MNRAS, 365, 1220
\bibitem{}Bell E.F. \& Kennicutt R.C. 2001, ApJ, 548, 681
\bibitem{}Boyce P.J., Minchin R.F., Kilborn V.A. et al. 2001, ApJ, 560, L127
\bibitem{}de Vaucouleurs, G.,  de Vaucouleurs, A., Corwin, A., Buta, R.J.,
	   Paturel, G., Fouq\'{u}e, P. 1991, Third Reference
	   Catalogue of Bright  Galaxies, New-York - Springer-Verlag
\bibitem{}de Vaucouleurs, G. 1975, in ``Galaxies and the Universe,'' Sandage A.,
      Sandage M., Kristian J. (eds.) Chicago, Univ. of Chicago Press, p. 557
\bibitem{}Dohm-Palmer R.C. et al. 2002, AJ, 123, 813
\bibitem{}Dolphin A.E. et al. 2003, AJ, 126, 187
\bibitem{}Gallagher J.S., Hunter D.A. \& Tutukov A.V. 1984, ApJ, 284, 544
\bibitem{}Gil de Paz, Madore B.F. \& Pevunova O. 2003, ApJS, 147, 29
\bibitem{}Giovanelli R. \& Haynes M.P. 1991, ARA\&A, 29, 499
\bibitem{}Hanish D.J., Meurer G.R., Ferguson H.C. et al. 2006, ApJ, 649, 150
\bibitem{}Hodge P.W. \& Kennicutt R.C. 1983, AJ, 88, 296
\bibitem{}Hoopes C.G., Walterbos R.A.M. \& Rand R.J. 1999, ApJ, 522, 669
\bibitem{}Huchtmeier W.K., Karachentsev I.D. \& Karachentseva V.E. 2003, A\&A, 401, 483
\bibitem{}Huchtmeier W.K., Karachentsev I.D., Karachentseva V.E. \& Ehle M.
    2000, A\&AS, 141, 469
\bibitem{}Hunter D.A. \& Elmegreen B.G. 2004, AJ, 128, 2170
\bibitem{}Hunter D.A. \& Gallagher J.S. 1992, ApJ, 391, L9
\bibitem{}James P.A., Shane N.S., Beckman J.E., et al. 2004, A\&A, 414, 23
\bibitem{}Karachentsev I.D. \& Kaisin S.S. 2007, AJ, 133, April
\bibitem{}Karachentsev I.D., Dolphin A.E., Tully R.B. 2006, AJ, 131, 1361
\bibitem{}Karachentsev I.D., Karachentseva V.E., Huchtmeier W.K., Makarov D.I.
       2004, AJ, 127, 2031
\bibitem{}Karachentsev I.D., Sharina M.E., Dolphin A.E. 2003, A\&A, 398, 467
\bibitem{}Karachentsev I.D. 1966, Astrofizika, 2, 81
\bibitem{}Kennicutt R.C. 1998, ApJ, 498, 541
\bibitem{}Kennicutt R.C. 1989, ApJ, 344, 685
\bibitem{}Kennicutt R.C., Edgar B.K., Hodge P.W. 1989, ApJ, 337, 761
\bibitem{}Makarova L.N., Karachentsev I.D. \& Georgiev T.B. 1997, Astron. Lett., 23, 435
\bibitem{}Martin D.C., Seibert M., Buat V. et al. 2005, ApJ, 619, 59
\bibitem{}McConnachie A.W., Arimoto N., Irwin M., Tolstoy E. 2006, MNRAS, 373, 715
\bibitem{}Miller B.W. \& Hodge P. 1994, ApJ, 427, 656
\bibitem{}Minchin R.F., Davies J., Disney M. et al. 2005, ApJ, 622, L21
\bibitem{}Nakamura O., Fukugita M. et al. 2004, AJ, 127, 2511
\bibitem{}Oke J.B. 1990, AJ, 99, 1621
\bibitem{}Papaderos P. et al. 1996, A\&AS, 120, 207
\bibitem{}Rekola R., Jerjen H. \& Flynn C. 2005, A\&A, 437, 823
\bibitem{}Sargent W.L.W. \& Searle L. 1970, ApJ, 162, L155
\bibitem{}Seth A.C., Dalcanton J.J. \& de Jong R.S. 2005, AJ, 129, 1331
\bibitem{}Schlegel D.J., Finkbeiner D.P. \& Davis M. 1998, ApJ, 500, 525
\bibitem{}Schneider S.E. 1985, ApJ, 288, L33
\bibitem{}Skillman E.D. 2005, New Astronomy Review, 49, 453
\bibitem{}Spergel D.N. et al. 2003, ApJS, 148, 175
\bibitem{}Stinson G.S., Dalcanton J.,J., Quinn T. et al. 2007, astro-ph/0705.4494
\bibitem{}Strickland D.K., Heckman T.M., Colbert E.J.M. et al. 2004, ApJ, 606, 829
\bibitem{}Taylor E.N. \& Webster R.L. 2005, ApJ, 634, 1067
\bibitem{}Telles E. \& Maddox S. 2000, MNRAS, 311, 307
\bibitem{}Tonry J.L. et al. 2001, ApJ, 546, 681
\bibitem{}Tully R.B., Shaya E.J. \& Pierce M.J. 1992, ApJS, 80, 479
\bibitem{}Tully R.B. et al. 2007, in prepapation
\bibitem{}Tully R.B. 1988, Nearby Galaxy Catalog, Cambridge Univ. Press
\bibitem{}Tutukov A.V. 2006, Astronomy Reports, 50, 526
\bibitem{}van Zee L. 2000, AJ, 119, 2757
\bibitem{}Vennik J. 1984, Tartu Astron. Obs. Publ. 73, 1
\bibitem{}Verheijen M.A.W. 2001, ApJ, 563, 694
\bibitem{}Weidner C. \& Kroupa P. 2005, ApJ, 625, 754
\bibitem{}Young J.S., Allen L., Kenney J.D. \& Rownd B. 1996, AJ, 112, 1903
\bibitem{}Young L.M., Skillman E.D., Weisz D.R. \& Dolphin A.E. 2007, ApJ, 659, 331
\end{thebibliography}
